\documentclass[11pt]{article}
\usepackage[utf8]{inputenc}
\usepackage{amsmath}
\usepackage{amsfonts}
\usepackage{fullpage,times}
\usepackage{amssymb}
\usepackage{graphicx}
\begin{document}
\title{\textbf{Introduction to Dynamic Unary Encoding}}
\date{\today}
\author{Ernst D. Berg  \\ ernst@eberg.us  \\ Turlock, CA.}
\maketitle


\begin{abstract} Dynamic Unary Encoding takes Unary Encoding to the next level. Every $n$\textit{-bit} binary string is an encoding of dynamic unary and every $n$\textit{-bit} binary string is encodable by dynamic unary. By utilizing both forms of unary code and a single bit of parity information dynamic unary encoding partitions $2^n$ non-negative integers into $n$ sets of disjoint cycles of $n$\textit{-bit} elements. These cycles have been employed as virtual data sets, binary transforms and as a mathematical object. Characterization of both the cycles and of the cycle spectrum is given. Examples of encoding and decoding algorithms are given. 
Examples of other constructs utilizing the principles of dynamic unary encoding are presented. The cycle as a  mathematical object is demonstrated. 
\end{abstract}

\section{Introduction}
Dynamic Unary Encoding (DUE), in the form of a binary data encoder, was discovered by this author in January of 2010 during an effort to satisfy a data compression challenge through experimental binary encoding schemes involving iterative functions. Although the effort to find a solution to that challenge continues; DUE is a discovery of importance. Effort is made to present DUE in an introductory format leading to suggestions of applications in higher mathematics and computer science. It has not escaped observation that Dynamic Unary has applications beyond data encoding.
     
\subsection{Overview} 
 The history and development of Dynamic Unary Encoding is a reflection of my search for innovative binary encoding resulting in four categories of algorithms developed through experimentation and observation. 
 
 The first category of encoder developed is introduced as an iterative function exporting the parity of the first bit, in the binary string being encoded, as the transform. An example of this category of algorithm is presented in 5.1.
 
  The second category includes binary string constructions and deconstructions exploring the relationship of parity information and unary code to binary string structure. An example is presented in 5.2.  
 
 The third category communicates the parity structure of the source string via the terminus parity of the unary code. Through iteration of encoding or decoding it was observed that for all $n$\textit{-bit} binary strings DUE generated a cycle of $n$\textit{-bit} elements.  
 
 For the fourth category the cycles of DUE as a complete orbit are successfully utilized as a mathematical object; a type of dynamic integer or Quantum Numbers perhaps? A non-DUE cycle is generated as the example of this category.
 
\subsection{Unary code} 
 Unary Code\cite{unary1} is an entropy code that represents a non-negative integer. Unary code has the form of a body of zero or more bits of one parity and a single bit terminus of the opposite parity.
\begin{itemize} \item Example: Integer value four can be represented in unary code as both 0111 and 1000\footnote{Binary\cite{binary} data are presented in the Western Positional Notation of right to left.} .
\end{itemize}	
A finite length binary string may contain more than one unary coded integer. Each integer is delimited by the terminus parity.
\begin{itemize} \item Example: In the binary string ``0111011110110100" unary codes with a terminus of the 0 parity represent 
integers  $\left\lbrace 4,5,3,2,1,1 \right\rbrace$. 
\end{itemize}
 
\subsection{Parity structure of binary strings and the Parity Reference } 
 Binary strings have the quality of being dividable into segments of same parity bits in an alternating parity pattern forming the parity structure of the binary string. 
 \begin{itemize}
\item Example: For \textbf{``} 0111011110110100 \textbf{"} same parity bit segments are  $\left\lbrace0,111,0,1111,0,11,0,1,00\right\rbrace $.
\end{itemize}
 This parity structure is communicated with a single bit by knowing the parity of a single bit at a specific bit position. The term for this is the Parity Reference.  The term Parity Reference incorporates both the physical location of that bit and the information of that bit's parity state. The Parity Reference may be internal to the encoding or can be external.  Bit position notation follows conventional representation of $b_{0}$ (first bit) through $b_{n-1}$ for some finite length binary string. A finite length binary string is also referenced as an $n$\textit{-bit}. 
 
\section{Encoding and Decoding}

 The Cycle and Mathematical Object aspects of Dynamic Unary Encoding are presented first followed by two encoding constructs utilizing external Parity Reference constructs. Dynamic Unary Encoding utilizes Unary Code and Parity Reference  information for all constructs.  

\subsection{Encoding}First, encoding parses a finite length binary string into same parity segments. The terminus parity of the unary code is determined by the parity of the bit at the Parity Reference bit position. Once the binary string parity structure is exported as terminus parity a simple bijection of segment length and unary code exists.  The lengths of those segments are then represented in a new string by unary codes. An example encode algorithm is presented in figure 1.
\begin{itemize}
\item Let $\sigma$ represent encoding. 
\item Let our source string be \begin{small}``0100100001100101011011000110110001101111"\end{small} ``Hello" in ASCII code. 
\item Let the Parity Reference =  $b_{0}$ making the unary code terminus = 1.
\end{itemize}
\begin{tabbing}
Same parity segments        \= 0 1 00 1 0000 11 00 1 0 1 0 11 0 11 000 11 0 11 000 11 0 1111\\
Unary Encoded \>  1 1 10 1 1000 10 10 1 1 1 1 10 1 10 100 10 1 10 100 10 1 1000 
\end{tabbing}
\begin{small}\textbf{ $\sigma(0100100001100101011011000110110001101111)_{0} =  (1110110001010111110110100101101001011000)_{1}$}
\end{small}
\subsection{Decoding}
First, decoding determines the terminus parity of the unary code used to encode from the $n$\textit{th} ($b_{n-1}$) bit of the string. Then the string is parsed into unary codes by terminus boundaries. Same parity segments, of the length the unary code communicates, are then written in correct parity order using the Parity Reference information. An example decode algorithm is presented in figure 2.
\begin{small}
\begin{itemize}
\item Let \textbf{-}$\sigma$ represent decoding.
\item Let our source string be ``1110110001010111110110100101101001011000" 
\item Let our unary code terminus be 1 and our Parity Reference = $b_{0}$. Terminus information is read at $b_{n-1}$.
\end{itemize}
\begin{tabbing}
Parse the string into unary codes     \= 1 1 10 1 1000 10 10 1 1 1 1 10 1 10 100 10 1 10 100 10 1 1000\\
Same parity bit segments \> 0 1 00 1 0000 11 00 1 0 1 0 11 0 11 000 11 0 11 000 11 0 1111
\end{tabbing}
\end{small}
\begin{small} \textbf{-$\sigma(1110110001010111110110100101101001011000)_{1} = (0100100001100101011011000110110001101111)_{0}$}
\end{small}

\section{Cycles and Cycle Spectrum} 

Every $n$\textit{-bit} binary string is an encoding of dynamic unary and every $n$\textit{-bit} binary string is encodable by dynamic unary. Through experimentation and observation iteration of encoding or decoding resulted in a cycle of $k$ elements of length $n$ bits. By utilizing both forms of unary code and the parity structure information of the source string dynamic unary partitions $2^n$ non-negative integers into disjoint cycles\cite{cycle1}\cite{cycle2} of $n$\textit{-bit} elements creating a dynamic data type. The spectrum of these cycles is an infinite progression of sets\cite{sets} of cycles that have in common relationships to powers of two.\footnote{A complete listing of the cycles for string lengths one through eight is presented in table one.} The quality of the cycles being disjoint was first determined by observations of the cycles of the shorter length strings which are easy to generate by hand. The disjoint quality of the cycles was also observed with a C-Language program generating the cycles of thirty two bit elements, as a data set, of all $2^{32}$ non-negative integers. Encoding is further defined as generating the next element in cycle and decoding is further defined as generating the previous element in cycle. 
\begin{small}
\begin{center}
$S_{0}$ = $\left\lbrace 0000\right\rbrace $  $n$ = 4 $k$ = 8  PRef = $b_{0}$ $S_{i+1} =  \sigma(S_{0})_{i}$ or \textbf{-}$\sigma(S_{0})_{i} $   for $0 \geq i < k$  $S_{k} = S_{0}$ \\
$ \sigma(0000)$ = ( 0111 1100 0101 1111 1000 0011 1010 0000 ) = ( 7 12 5 15 8 3 10 0 )

\textbf{-}$\sigma(0000)$ = ( 1010 0011 1000 1111 0101 1100 0111 0000 ) = (  10 3 8 15 5 12 7  0  )
$S_{0} \mapsto S_{1} \mapsto S_{2} \mapsto S_{3} \mapsto S_{4} \mapsto S_{5} \mapsto S_{6} \mapsto S_{7} = S_{0}$ \\ Example cycle in both the encode direction and the decode direction of a \textit{$4$-bit} string.
\end{center} 
\end{small} 
  The number of cycles for a $n$\textit{-bit} binary string and a specific Parity Reference bit position is determined by $x$ equals the number of elements in a cycle and $2^n$ the number of non-negative integers such that $cycles = 2^n/x$. That there are $n$ choices for a Parity Reference means we also have $n$ sets of cycles for each $n$\textit{-bit} string.

\begin{small}
\begin{itemize} 
\item From observations of the cycle data, cycle spectrum can be separated into three categories. 
\item String length $(2^n)^{b_{0}}$ has ($2^{1+n}$) cycles of ($2^{n}$) elements for $ n = 0 $ .\footnote{In researching the Mathematics of Cycles it was observed that a cycling of a single element is not generally considered a Cycle;\cite{cycle1} however, what applies to a string of one bit, in DUE, applies to all strings and is therefore included as such.} 
\item Binary string lengths that are a power of two, greater than one, and Parity Reference of $b_{0}$ have been observed as having an element count of ($\displaystyle{2^{1+n}}$) and cycle count of ($\displaystyle{2^{2^n}/2^{1+n}}$) for $n\in\{1,2,3,\dots\}$. 
\item All other lengths and Parity Reference combinations may be defined as $(2^{1+\lfloor \log_2 n \rfloor})$ elements \footnote{$(2^{1+\lfloor \log_2 n \rfloor})_{n\in\{1,2,3,\ldots\}}$ = $\left\lbrace 2,4,4,8,8,8,8,16,16,16,16,16,16,16,16,32,32,32,... \right\rbrace $} and ($2^{1+n}/2^{1+\lfloor \log_2 n \rfloor})$ cycles for $n\in\{1,2,3,\ldots\}$. 
\end{itemize}

\end{small}
\begin{flushleft}
\section{The Cycle as a mathematical object} 

\subsection{Introduction} By utilizing the complete orbit, application as a mathematical object is realized.  As a demonstration a non-DUE cycle is presented. The concepts of ``Cycle-Of" ($\sigma_{Of}$) and ``Cycle-On" ($\sigma_{On}$) are now introduced to differentiate types of cycles. Cycle-Of is defined as the dynamic unary cycle of an element. Cycle-On is defined as applying Cycle-Of to an element. In the following example a function utilizing the orbits of three integers and the logical operation exclusive-or\cite{xor}, a ($\sigma_{On}$) cycle of thirty two elements is generated. That for the element length of sixteen bits and a Parity Reference $b_{0}$ the number of ($\sigma_{Of}$) elements is thirty two and the number of elements in the ($\sigma_{On}$) is also thirty two. This is not always the case.  For example change the Parity Reference to any other bit position and the number of elements in the ($\sigma_{Of}$) cycle is sixteen. Curiously the number of elements in the ($\sigma_{On}$) orbit is still thirty two. Thus for Parity Reference $b_{0}$ only one ($\sigma_{Of}$) orbit occurs and for all other  Parity References two ($\sigma_{Of}$) orbits are required to generate a complete ($\sigma_{On}$) orbit. The halting condition of both the Cycle-Of and the Cycle-On is when the element generated equals the starting element. The reason only one ($\sigma_{Of}$) orbit occurs for string lengths of powers of two and a Parity Reference of $b_{0}$ is seen in the element structure of the cycle in that the elements of those cycles can be divided into two sets where the members of each are parity inverses of the other.  
\begin{itemize}
\item An example of length four bits $\sigma_{Of}$(0000) = (0111 1100 0101 1111 1000 0011 1010 0000 )
\end{itemize}
\center ---------------------------\\

For the following initial conditions this function generates a ($\sigma_{On}$) orbit for the element $S_{0} = 2014$ .  
\end{flushleft}
\begin{itemize}
\begin{small}
\item Let the length of string be sixteen bits.
\item Let the Parity Reference be $b_{0}$.
\item Let $0 \leq j < 32$ \space and \space $0 \leq i < 32$ \space where \space  $S_{32} = S_{0}$,\space \space $X_{32}=X_{0}$,\space \space $Y_{32}=Y_{0}$ \space and \space $Z_{32}=Z_{0}$ 
\item Let $X_{0}=1$,\space $Y_{0}=99$,\space $Z_{0}=6408$\space and \space$S_{0}=2014$
\item Let $S_{j+1} = S_{j} \oplus \sigma(X)_{i} \oplus \sigma(Y)_{i} \oplus \sigma(Z)_{i} $  
 \begin{tabbing} The \emph{Cycle-On} cycle \space  \textbf{$:$} \space \textbf{$\sigma_{On}(2014)$} = ( \space \= 28158 19761 64921 60058 30232 23332 8057 63754 \\ \> 27712 19536 951 60323 34882 23123 57674 2015 \\ \> 37376 19760 615 60059 35302 23333 57479 63755 \\ \> 37822 19537 64585 60322 30652 23122 7860 2014 \space ) \end{tabbing}

\end{small}
\end{itemize} 
\begin{flushleft}

\indent{ } If we examine the mapping it is clear that the next element in the ($\sigma_{On}$) cycle is not a result of Dynamic Unary Encoding. $0000011111011110 \mapsto 0110110111111110$.  The value 2,014 is mapped to the value 28,158. 
A quality of this ($\sigma_{On}$) orbit is that by knowing one element in the cycle and it's index the original element ($S_{0}$) can be retrieved.
 
\indent{ } This example utilizes cycles of the same type, direction of spin and element count for simplicity of demonstration however constructs are not limited to same type of cycle, same number of cycles in a function, same spin direction or the cycles being of the same \emph{$k$} size. 
 
\subsection{Observations}It has been observed that the cycles with element lengths, greater than one, that are a power of two of Parity Reference $b_{0}$ and powers of two plus one of Parity Reference $b_{1}$ have elements, when considered to be unsigned integers, that sum to the same value. 
\begin{itemize}
\begin{small} 
\item Let $\alpha$ represent binary strings length $2^n$ and $\beta$ represent binary strings length $2^n +1$ for  $0 < n \leq\infty$ \end{small} 
\item $\sum ^{2(2^n)} _{i=1} \sigma(\alpha)_{i} ^{b_{0}}$ = $(2^{2^n}-1)2^n $ \begin{small}$ \textbf{:}$ \space $\left\lbrace  \sum \sigma({\alpha}) |  6,60,2040,1048560,137438953440, \dots\right\rbrace $\end{small}
\item $\sum ^{(2^{1+\lfloor \log_2 (2^n+1 )\rfloor})} _{i=1} \sigma(\beta)_{i} ^{b_{1}}$ = $(2^{2^n+1}-1)2^n $   \begin{small}$ \textbf{:}$ \space $\left\lbrace \sum \sigma({\beta}) |  14,124,4088,2097136,274877906912, \dots\right\rbrace $\end{small}
\end{itemize}
\indent{ }Other $n$\textit{-bit} and Parity Reference combinations have been observed as those cycles having more than a single sum suggesting that those cycles may be grouped by a conserved quantity quality. 

\section{Alternate algorithms}Dynamic Unary has two components, the Unary Code and the Parity Reference information. Each part can be utilized in constructs separately. Two encoding schemes are presented as examples. 
\subsection{Drop-T}The Drop-T algorithm exports the parity at Parity Reference $b_{0}$ as the encoding transform. The source $n$\textit{-bit} is then encoded in fixed parity terminus unary codes. Each iteration of the encoding then ``drops" the $n$\textit{th} bit of the newly unary encoded $n$\textit{-bit} source string before the next iteration. Each iteration reduces the source $n$\textit{-bit} length by one bit and increases output $n$\textit{-bit} length by one bit until all source bits are processed. 

\indent{ } Decoding requires that the terminus parity be known. Decoding reads each bit sequentially from the source $n$\textit{-bit} providing the parity information needed to reconstruct the parity structure of the $n$\textit{-bit} binary string of each iteration. Because the construct of encoding is the collection of all the parities of each iteration, Parity Reference is considered external to this $n$\textit{-bit} being constructed. Decoding starts with a single bit unary code and processes it's parity structure then adds a new terminus bit to the end of that $n$\textit{-bit} increasing the length of the $n$\textit{-bit} string by one bit. This iterative process completes when all the source $n$\textit{-bit} bits are used.

\begin{itemize} \begin{small}
\item Let our source be $\left\lbrace011\right\rbrace$  and our Parity Reference be $b_{0}$. Our fixed parity terminus is 1.
\item The encoded string then is all the parities sampled at $b_{0}$ \\ $\left\lbrace 011\right\rbrace \mapsto \left\lbrace 110\right\rbrace$ , $\left\lbrace 10\right\rbrace \mapsto \left\lbrace 11\right\rbrace$ , $\left\lbrace 1\right\rbrace \mapsto \left\lbrace 1\right\rbrace $  so $\left\lbrace 011\right\rbrace \mapsto \left\lbrace 101\right\rbrace$ 

\indent{ } It was hoped that Drop-T would be a single bijective cycle for all $n$\textit{-bit} elements representing all $2^n$ non-negative integers however it was discovered that some elements maped to themselves.
\item With a terminus parity of 1 and a string of ``000" the element maps to itself \\ $\left\lbrace 000\right\rbrace \mapsto\left\lbrace 100\right\rbrace , \left\lbrace 00\right\rbrace \mapsto\left\lbrace 10\right\rbrace , \left\lbrace 0\right\rbrace \mapsto\left\lbrace 1\right\rbrace $ so $\left\lbrace 000\right\rbrace \mapsto\left\lbrace 000\right\rbrace $ 

\indent{ } Therefore there are more than one cycle for each $n$\textit{-bit}.
\item Let our Parity Reference be $b_{0}$. Our fixed parity terminus is 0.
\item For length one the cycles are $\left\lbrace  (0),(1)  \right\rbrace $
\item For length two the cycles are $\left\lbrace \right.$( 00 01 10 ),( 11 ) $\left. \right\rbrace$  
\item For length three the cycles are  $\left\lbrace \right.$( 000 011 110 ),( 001 100 010 ),( 101 ),( 111 )$\left. \right\rbrace$
\end{small}\end{itemize} 
\end{flushleft}
\begin{flushleft}
\subsection{Binary string construction and deconstruction algorithms}During this phase of development several construction and deconstruction concepts were explored in the hopes of finding some ``data encoding magic."  Imagination was key during this phase. The basic constructs of this category process unary code separately from Parity Reference information. Several variations of this type of encoding were explored resulting in unique encoding schemes. For this example, any string length $n$, two strings of length $n+1$ can be generated. The Parity Reference is an external and fixed parity. 
\end{flushleft}
\begin{itemize}
\item Given an arbitrary string ``011" then adding terminus' 0+011 and 1+011 then decoding for a fixed Parity Reference of 0 we get $\left\lbrace 1000\right\rbrace $ and $\left\lbrace 0010\right\rbrace $ because it is assumed the $n$\textit{th} bit is the terminus parity. Therefore it is possible to generate an eight bit string from three bits of information. 
\item So for this decode $\left\lbrace 011 \right\rbrace \mapsto \left\lbrace 10000010 \right\rbrace $
\end{itemize}
\begin{flushleft}
 
\section*{Related work}
 The cycle structure was first observed by Gustavus J. Simmons\cite{simmons} and is presented in the paper \emph{Parity Encoding of Binary Sequences}. Simmons applies an exclusive-or summation algorithm; a \emph{``mod 2 sum of the summands"},
 on binary strings which have in common a finite length and a parity of 1 in the most significant bit position. The cycles generated prove to be a subset of those generated by DUE for the conditions of the decode direction with the Parity Reference bit position of $b_{n-1} $ when that bit is parity 1.\footnote{Table One presents the encode data in left to right order. Decode direction equals right to left order.} It is interesting that two different algorithms generate the cycle structure and raises the question of what do exclusive-or and DUE have mathematically in common?  
\section*{Conclusion}Dynamic Unary Encoding introduces new choices of how binary data is represented and processed. DUE emerged out of efforts to discover innovative encoding methods. DUE has been applied as a bijective transform changing one binary data file into millions of different files. DUE was an important part of a surjective data encoder where many choices of what binary code would represent a source $n$\textit{-bit} string were generated. As a tool of deconstruction or construction binary data is processed bit by bit. Utilized as a mathematical object, the forms a data can take is limited only by imagination because anything that can be done and undone to a data can be a part of some encoding algorithm utilizing DUE. As a virtual data set dynamic unary requires only one element in order to access all $k$ elements of a cycle. That quality also extends to simple encryption of data by simply cycling a data in it's orbit.

\indent{ } It is not difficult to envision applications of dynamic unary in categories such as the design of microprocessors, robotics, artificial intelligence, cryptography, logic and control and possibly physics. 

\indent{ } \emph{All in all} there is more to DUE than has been done. More to discover about Dynamic Unary and it's applications. The questions now are what needs to be done, how is it to be applied and what results will be realized. 

\indent{ }On a personal note, like Gottfried Leibniz\cite{leibniz} I too find the relationship of binary numbers and the \emph{I-Ching}\cite{Iching} interesting. Dynamic Unary introduces a quality that I believe Leibniz would find interesting and that is, as noted in the \emph{I-Ching} and \emph{Dynamic Unary Encoding}, that one element can change to another. Metaphorically speaking, a Gate of Wonders\footnote{Duyvendal's translation is recommended as the context of this comment. }\cite{tao} has been opened and the \emph{garden} that is DUE awaits.
   
\section*{Acknowledgments}
\indent{ } I wish to thank Gustavus Simmons for the paper \emph{Parity Encoding of Binary Sequences} which has provided both an independent validation of the cycle structure and a guide for this my first paper.
    
\indent{ } Thanks goes to Henning Makholm and mathstackexchange.com for assistance with the element progression equation.

\indent{ } J.J.L. Duyvendal's translation  of chapter one of the Tao Te Ching is unique among many and has been an inspiration in my efforts to understand the nature of information.. 
 
\indent{ } Also, thanks goes to Mark Nelson whose Million Digit Challenge has provided me with a trustworthy data-set and an incredibly difficult challenge. And to all those who have been a part of that experience through the years.
\section*{Dedication}
\indent{ } In memory of \emph{James Herbert Jones}; my step-father not forgotten, observing his early Sunday morning maths doodles with coffee was the genesis of my realization that there is a realm of creativity to Mathematics.

\begin{small}

\end{small}
\end{flushleft}
\begin{figure}[hbtp]
\centering
\includegraphics[scale=.70]{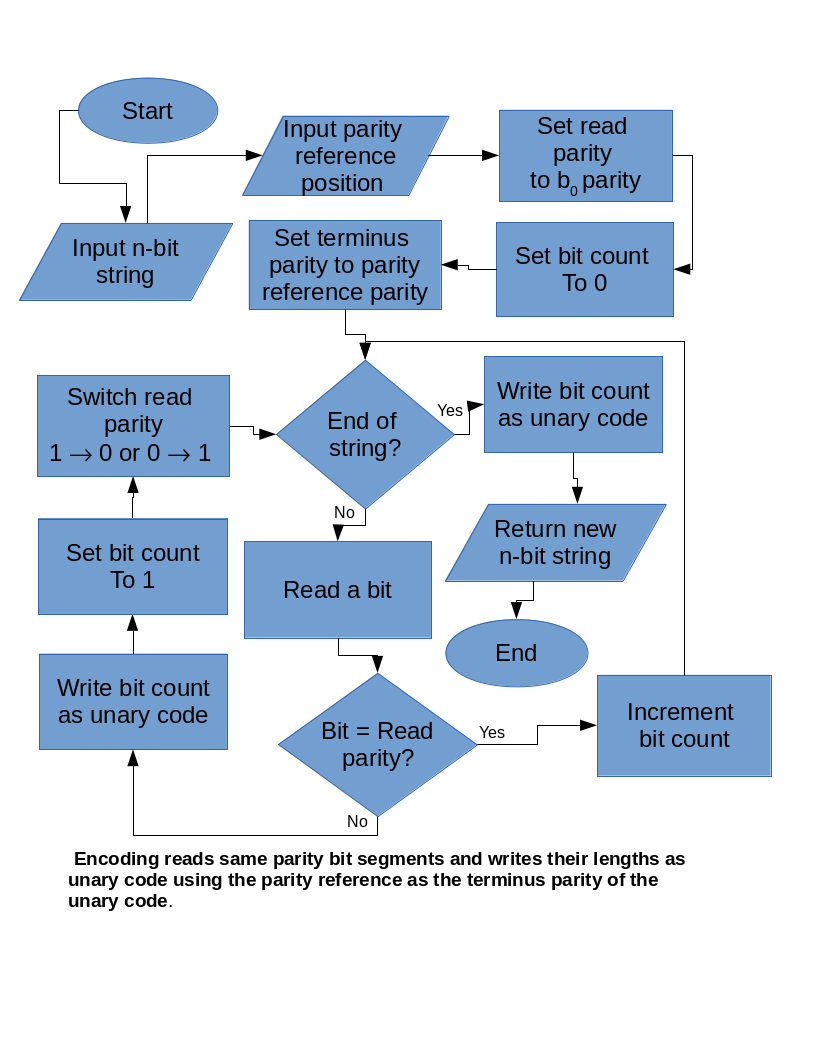}
\caption{Example encode algorithm}
\end{figure}
\begin{flushleft}
\begin{figure}[hbtp]
\includegraphics[scale=.70]{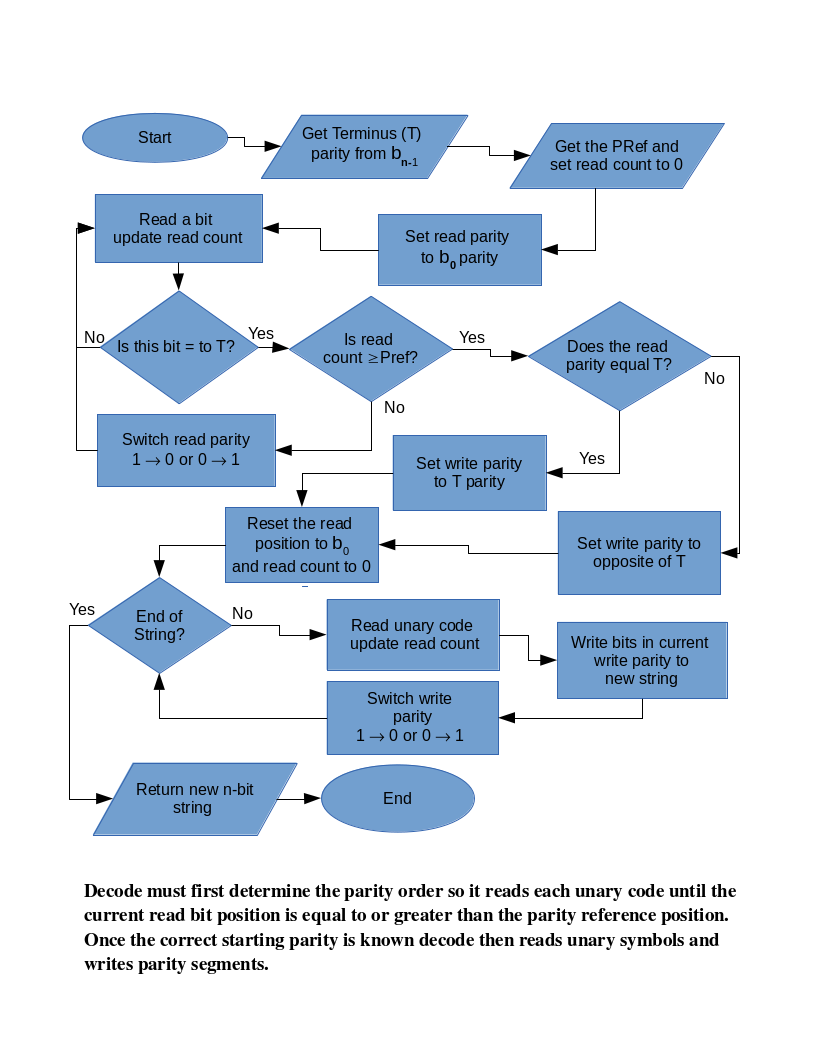}
\caption{Example decode algorithm}
\end{figure}
\end{flushleft}
\newpage
\begin{tiny}
\begin{center}

========================================================

\textbf{Length of string 1 bit.}  
 
(  0  ) (  1  )  \textbf{ Parity reference $b_{0}$ Number of Cycles 2 of 1 Element}

----------------

\textbf{Length of string 2 bits.}  

(  1  3  2  0  )  \textbf{ Parity reference $b_{0}$ Number of Cycles 1 of 4 Elements}

(  1  0  ) (  3  2  ) \textbf{ Parity reference $b_{1}$ Number of Cycles 2 of 2 Elements}

----------------

\textbf{Length of string 3 bits.}

 (  3  6  2  0  ) (  5  7  4  1  )
 \textbf{ Parity reference $b_{0}$ Number of Cycles 2 of 4 Elements}

 (  3  6  5  0  ) (  2  7  4  1  )
 \textbf{ Parity reference $b_{1}$ Number of Cycles 2 of 4 Elements}

 (  3  1  2  0  ) (  6  5  7  4  )
 \textbf{ Parity reference $b_{2}$ Number of Cycles 2 of 4 Elements}

----------------

\textbf{Length of string 4 bits.}

 (  7  12  5  15  8  3  10  0  ) (  9  13  11  14  6  2  4  1  )
 \textbf{ Parity reference $b_{0}$ Number of Cycles 2 of 8 Elements}

 (  7  12  5  0  ) (  6  13  4  1  ) (  11  14  9  2  ) (  10  15  8  3  )
 \textbf{ Parity reference $b_{1}$ Number of Cycles 4 of 4 Elements}

 (  7  12  10  0  ) (  6  13  11  1  ) (  4  14  9  2  ) (  5  15  8  3  )
 \textbf{ Parity reference $b_{2}$ Number of Cycles 4 of 4 Elements}

 (  7  3  5  0  ) (  6  2  4  1  ) (  12  10  15  8  ) (  13  11  14  9  )
 \textbf{ Parity reference $b_{3}$ Number of Cycles 4 of 4 Elements}

----------------

\textbf{Length of string 5 bits.}

 (  15  24  11  30  14  6  10  0  ) (  17  25  21  31  16  7  20  1  ) (  12  5  23  28  13  27  22  2  ) (  18  4  9  29  19  26  8  3  )

 \textbf{ Parity reference $b_{0}$ Number of Cycles 4 of 8 Elements}

 (  15  24  11  30  17  6  21  0  ) (  14  25  10  31  16  7  20  1  ) (  19  26  23  28  13  4  9  2  ) (  18  27  22  29  12  5  8  3  )

 \textbf{ Parity reference $b_{1}$ Number of Cycles 4 of 8 Elements}

 (  15  24  11  1  14  25  10  0  ) (  12  26  8  3  13  27  9  2  ) (  22  29  19  5  23  28  18  4  ) (  21  31  16  7  20  30  17  6  )

 \textbf{ Parity reference $b_{2}$ Number of Cycles 4 of 8 Elements}

 (  15  24  20  1  14  25  21  0  ) (  12  26  23  3  13  27  22  2  ) (  9  29  19  5  8  28  18  4  ) (  10  31  16  7  11  30  17  6  )

 \textbf{ Parity reference $b_{3}$ Number of Cycles 4 of 8 Elements}

 (  15  7  11  1  14  6  10  0  ) (  12  5  8  3  13  4  9  2  ) (  24  20  30  17  25  21  31  16  ) (  27  22  29  19  26  23  28  18  )

 \textbf{ Parity reference $b_{4}$ Number of Cycles 4 of 8 Elements}

----------------

\textbf{Length of string 6 bits.}

 (  31  48  23  60  29  51  42  0  ) (  33  49  41  61  35  50  20  1  ) (  28  13  43  62  30  14  22  2  ) (  34  12  21  63  32  15  40  3  )

 (  25  53  47  56  27  54  18  4  ) (  39  52  17  57  37  55  44  5  ) (  26  8  19  58  24  11  46  6  ) (  36  9  45  59  38  10  16  7  )

 \textbf{ Parity reference $b_{0}$ Number of Cycles 8 of 8 Elements}

 (  31  48  23  60  29  12  21  0  ) (  30  49  22  61  28  13  20  1  ) (  35  50  43  62  33  14  41  2  ) (  34  51  42  63  32  15  40  3  )

 (  25  10  47  56  27  54  45  4  ) (  24  11  46  57  26  55  44  5  ) (  37  8  19  58  39  52  17  6  ) (  36  9  18  59  38  53  16  7  )

 \textbf{ Parity reference $b_{1}$ Number of Cycles 8 of 8 Elements}

 (  31  48  23  60  34  12  42  0  ) (  30  49  22  61  35  13  43  1  ) (  28  50  20  62  33  14  41  2  ) (  29  51  21  63  32  15  40  3  )

 (  38  53  47  56  27  9  18  4  ) (  39  52  46  57  26  8  19  5  ) (  37  55  44  58  24  11  17  6  ) (  36  54  45  59  25  10  16  7  )

 \textbf{ Parity reference $b_{2}$ Number of Cycles 8 of 8 Elements}

 (  31  48  23  3  29  51  21  0  ) (  30  49  22  2  28  50  20  1  ) (  25  53  16  7  27  54  18  4  ) (  24  52  17  6  26  55  19  5  )

 (  44  58  39  11  46  57  37  8  ) (  45  59  38  10  47  56  36  9  ) (  42  63  32  15  40  60  34  12  ) (  43  62  33  14  41  61  35  13  )

 \textbf{ Parity reference $b_{3}$ Number of Cycles 8 of 8 Elements}

 (  31  48  40  3  29  51  42  0  ) (  30  49  41  2  28  50  43  1  ) (  25  53  47  7  27  54  45  4  ) (  24  52  46  6  26  55  44  5  )

 (  19  58  39  11  17  57  37  8  ) (  18  59  38  10  16  56  36  9  ) (  21  63  32  15  23  60  34  12  ) (  20  62  33  14  22  61  35  13  )

 \textbf{ Parity reference $b_{4}$ Number of Cycles 8 of 8 Elements}

 (  31  15  23  3  29  12  21  0  ) (  30  14  22  2  28  13  20  1  ) (  25  10  16  7  27  9  18  4  ) (  24  11  17  6  26  8  19  5  )

 (  48  40  60  34  51  42  63  32  ) (  49  41  61  35  50  43  62  33  ) (  54  45  59  38  53  47  56  36  ) (  55  44  58  39  52  46  57  37  )

 \textbf{ Parity reference $b_{5}$ Number of Cycles 8 of 8 Elements}

----------------

\textbf{Length of string 7 bits.}

 (  63  96  47  120  59  102  42  0  ) (  65  97  81  121  69  103  84  1  ) (  60  29  83  122  56  27  86  2  ) (  66  28  45  123  70  26  40  3  )

 (  57  101  87  124  61  99  82  4  ) (  71  100  41  125  67  98  44  5  ) (  58  24  43  126  62  30  46  6  ) (  68  25  85  127  64  31  80  7  )

 (  51  106  32  15  72  19  90  8  ) (  77  107  94  14  54  18  36  9  ) (  48  23  92  13  75  110  38  10  ) (  78  22  34  12  53  111  88  11  )

 (  39  116  49  105  93  115  74  16  ) (  89  117  79  104  35  114  52  17  ) (  33  113  73  109  91  118  50  20  ) (  95  112  55  108  37  119  76  21  )

 \textbf{ Parity reference $b_{0}$ Number of Cycles 16 of 8 Elements}

 (  63  96  47  120  59  102  85  0  ) (  62  97  46  121  58  103  84  1  ) (  67  98  83  122  71  100  41  2  ) (  66  99  82  123  70  101  40  3  )

 (  57  26  87  124  61  28  45  4  ) (  56  27  86  125  60  29  44  5  ) (  69  24  43  126  65  30  81  6  ) (  68  25  42  127  64  31  80  7  )

 (  51  106  95  112  55  108  37  8  ) (  50  107  94  113  54  109  36  9  ) (  79  104  35  114  75  110  89  10  ) (  78  105  34  115  74  111  88  11  )

 (  53  16  39  116  49  22  93  12  ) (  52  17  38  117  48  23  92  13  ) (  73  18  91  118  77  20  33  14  ) (  72  19  90  119  76  21  32  15  )

 \textbf{ Parity reference $b_{1}$ Number of Cycles 16 of 8 Elements}

 (  63  96  47  120  59  25  42  0  ) (  62  97  46  121  58  24  43  1  ) (  60  98  44  122  56  27  41  2  ) (  61  99  45  123  57  26  40  3  )

 (  70  101  87  124  66  28  82  4  ) (  71  100  86  125  67  29  83  5  ) (  69  103  84  126  65  30  81  6  ) (  68  102  85  127  64  31  80  7  )

 (  51  21  95  112  55  108  90  8  ) (  50  20  94  113  54  109  91  9  ) (  48  23  92  114  52  110  89  10  ) (  49  22  93  115  53  111  88  11  )

 (  74  16  39  116  78  105  34  12  ) (  75  17  38  117  79  104  35  13  ) (  73  18  36  118  77  107  33  14  ) (  72  19  37  119  76  106  32  15  )

 \textbf{ Parity reference $b_{2}$ Number of Cycles 16 of 8 Elements}

 (  63  96  47  120  68  25  85  0  ) (  62  97  46  121  69  24  84  1  ) (  60  98  44  122  71  27  86  2  ) (  61  99  45  123  70  26  87  3  )

 (  57  101  40  124  66  28  82  4  ) (  56  100  41  125  67  29  83  5  ) (  58  103  43  126  65  30  81  6  ) (  59  102  42  127  64  31  80  7  )

 (  76  106  95  112  55  19  37  8  ) (  77  107  94  113  54  18  36  9  ) (  79  104  92  114  52  17  38  10  ) (  78  105  93  115  53  16  39  11  )

 (  74  111  88  116  49  22  34  12  ) (  75  110  89  117  48  23  35  13  ) (  73  109  91  118  50  20  33  14  ) (  72  108  90  119  51  21  32  15  )

 \textbf{ Parity reference $b_{3}$ Number of Cycles 16 of 8 Elements}

 (  63  96  47  7  59  102  42  0  ) (  62  97  46  6  58  103  43  1  ) (  60  98  44  5  56  100  41  2  ) (  61  99  45  4  57  101  40  3  )

 (  51  106  32  15  55  108  37  8  ) (  50  107  33  14  54  109  36  9  ) (  48  104  35  13  52  110  38  10  ) (  49  105  34  12  53  111  39  11  )

 (  88  116  78  22  93  115  74  16  ) (  89  117  79  23  92  114  75  17  ) (  91  118  77  20  94  113  73  18  ) (  90  119  76  21  95  112  72  19  )

 (  84  126  65  30  81  121  69  24  ) (  85  127  64  31  80  120  68  25  ) (  87  124  66  28  82  123  70  26  ) (  86  125  67  29  83  122  71  27  )

 \textbf{ Parity reference $b_{4}$ Number of Cycles 16 of 8 Elements}

 (  63  96  80  7  59  102  85  0  ) (  62  97  81  6  58  103  84  1  ) (  60  98  83  5  56  100  86  2  ) (  61  99  82  4  57  101  87  3  )

 (  51  106  95  15  55  108  90  8  ) (  50  107  94  14  54  109  91  9  ) (  48  104  92  13  52  110  89  10  ) (  49  105  93  12  53  111  88  11  )

 (  39  116  78  22  34  115  74  16  ) (  38  117  79  23  35  114  75  17  ) (  36  118  77  20  33  113  73  18  ) (  37  119  76  21  32  112  72  19  )

 (  43  126  65  30  46  121  69  24  ) (  42  127  64  31  47  120  68  25  ) (  40  124  66  28  45  123  70  26  ) (  41  125  67  29  44  122  71  27  )

 \textbf{ Parity reference $b_{5}$ Number of Cycles 16 of 8 Elements}

 (  63  31  47  7  59  25  42  0  ) (  62  30  46  6  58  24  43  1  ) (  60  29  44  5  56  27  41  2  ) (  61  28  45  4  57  26  40  3  )

 (  51  21  32  15  55  19  37  8  ) (  50  20  33  14  54  18  36  9  ) (  48  23  35  13  52  17  38  10  ) (  49  22  34  12  53  16  39  11  )

 (  96  80  120  68  102  85  127  64  ) (  97  81  121  69  103  84  126  65  ) (  99  82  123  70  101  87  124  66  ) (  98  83  122  71  100  86  125  67  )

 (  108  90  119  76  106  95  112  72  ) (  109  91  118  77  107  94  113  73  ) (  111  88  116  78  105  93  115  74  ) (  110  89  117  79  104  92  114  75  )

 \textbf{ Parity reference $b_{6}$ Number of Cycles 16 of 8 Elements}

----------------

\newpage
\textbf{Length of string 8 bits.}

 (  127  192  95  240  119  204  85  255  128  63  160  15  136  51  170  0  ) (  129  193  161  241  137  205  171  254  126  62  94  14  118  50  84  1  ) 
 
 (  124  61  163  242  116  49  169  253  131  194  92  13  139  206  86  2  ) (  130  60  93  243  138  48  87  252  125  195  162  12  117  207  168  3  )

 (  121  197  167  244  113  201  173  251  134  58  88  11  142  54  82  4  ) (  135  196  89  245  143  200  83  250  120  59  166  10  112  55  172  5  ) 
 
 (  122  56  91  246  114  52  81  249  133  199  164  9  141  203  174  6  ) (  132  57  165  247  140  53  175  248  123  198  90  8  115  202  80  7  )

 (  103  212  65  225  145  217  181  239  152  43  190  30  110  38  74  16  ) (  153  213  191  224  111  216  75  238  102  42  64  31  144  39  180  17  ) 
 
 (  100  41  189  227  146  36  73  237  155  214  66  28  109  219  182  18  ) (  154  40  67  226  108  37  183  236  101  215  188  29  147  218  72  19  )

 (  97  209  185  229  151  220  77  235  158  46  70  26  104  35  178  20  ) (  159  208  71  228  105  221  179  234  96  47  184  27  150  34  76  21  ) 
 
 (  98  44  69  231  148  33  177  233  157  211  186  24  107  222  78  22  ) (  156  45  187  230  106  32  79  232  99  210  68  25  149  223  176  23  )

 \textbf{ Parity reference $b_{0}$ Number of Cycles 16 of 16 Elements}

 (  127  192  95  240  119  204  85  0  ) (  126  193  94  241  118  205  84  1  )  (  131  194  163  242  139  206  169  2  ) (  130  195  162  243  138  207  168  3  )

 (  121  58  167  244  113  54  173  4  ) (  120  59  166  245  112  55  172  5  )  (  133  56  91  246  141  52  81  6  ) (  132  57  90  247  140  53  80  7  )

 (  115  202  175  248  123  198  165  8  ) (  114  203  174  249  122  199  164  9  ) (  143  200  83  250  135  196  89  10  ) (  142  201  82  251  134  197  88  11  )

 (  117  48  87  252  125  60  93  12  ) (  116  49  86  253  124  61  92  13  ) (  137  50  171  254  129  62  161  14  ) (  136  51  170  255  128  63  160  15  )

 (  103  212  65  30  145  38  181  16  ) (  102  213  64  31  144  39  180  17  ) (  155  214  189  28  109  36  73  18  ) (  154  215  188  29  108  37  72  19  )

 (  97  46  185  26  151  220  77  20  ) (  96  47  184  27  150  221  76  21  ) (  157  44  69  24  107  222  177  22  ) (  156  45  68  25  106  223  176  23  )

 (  79  232  99  210  187  230  149  32  ) (  78  233  98  211  186  231  148  33  ) (  179  234  159  208  71  228  105  34  ) (  178  235  158  209  70  229  104  35  )

 (  67  226  147  218  183  236  101  40  ) (  66  227  146  219  182  237  100  41  ) (  191  224  111  216  75  238  153  42  ) (  190  225  110  217  74  239  152  43  )

 \textbf{ Parity reference $b_{1}$ Number of Cycles 32 of 8 Elements}

 (  127  192  95  240  119  204  170  0  ) (  126  193  94  241  118  205  171  1  ) (  124  194  92  242  116  206  169  2  ) (  125  195  93  243  117  207  168  3  )

 (  134  197  167  244  142  201  82  4  ) (  135  196  166  245  143  200  83  5  ) (  133  199  164  246  141  203  81  6  ) (  132  198  165  247  140  202  80  7  )

 (  115  53  175  248  123  57  90  8  ) (  114  52  174  249  122  56  91  9  ) (  112  55  172  250  120  59  89  10  ) (  113  54  173  251  121  58  88  11  )

 (  138  48  87  252  130  60  162  12  ) (  139  49  86  253  131  61  163  13  ) (  137  50  84  254  129  62  161  14  ) (  136  51  85  255  128  63  160  15  )

 (  103  212  190  225  110  217  74  16  ) (  102  213  191  224  111  216  75  17  ) (  100  214  189  227  109  219  73  18  ) (  101  215  188  226  108  218  72  19  )

 (  158  209  70  229  151  220  178  20  ) (  159  208  71  228  150  221  179  21  ) (  157  211  69  231  148  222  177  22  ) (  156  210  68  230  149  223  176  23  )

 (  107  33  78  233  98  44  186  24  ) (  106  32  79  232  99  45  187  25  ) (  104  35  77  235  97  46  185  26  ) (  105  34  76  234  96  47  184  27  )

 (  146  36  182  237  155  41  66  28  ) (  147  37  183  236  154  40  67  29  ) (  145  38  181  239  152  43  65  30  ) (  144  39  180  238  153  42  64  31  )

 \textbf{ Parity reference $b_{2}$ Number of Cycles 32 of 8 Elements}

 (  127  192  95  240  119  51  85  0  ) (  126  193  94  241  118  50  84  1  ) (  124  194  92  242  116  49  86  2  ) (  125  195  93  243  117  48  87  3  )

 (  121  197  88  244  113  54  82  4  ) (  120  196  89  245  112  55  83  5  ) (  122  199  91  246  114  52  81  6  ) (  123  198  90  247  115  53  80  7  )

 (  140  202  175  248  132  57  165  8  ) (  141  203  174  249  133  56  164  9  ) (  143  200  172  250  135  59  166  10  ) (  142  201  173  251  134  58  167  11  )

 (  138  207  168  252  130  60  162  12  ) (  139  206  169  253  131  61  163  13  ) (  137  205  171  254  129  62  161  14  ) (  136  204  170  255  128  63  160  15  )

 (  103  43  190  225  110  217  181  16  ) (  102  42  191  224  111  216  180  17  ) (  100  41  189  227  109  219  182  18  ) (  101  40  188  226  108  218  183  19  )

 (  97  46  185  229  104  220  178  20  ) (  96  47  184  228  105  221  179  21  ) (  98  44  186  231  107  222  177  22  ) (  99  45  187  230  106  223  176  23  )

 (  148  33  78  233  157  211  69  24  ) (  149  32  79  232  156  210  68  25  ) (  151  35  77  235  158  209  70  26  ) (  150  34  76  234  159  208  71  27  )

 (  146  36  73  237  155  214  66  28  ) (  147  37  72  236  154  215  67  29  ) (  145  38  74  239  152  212  65  30  ) (  144  39  75  238  153  213  64  31  )

 \textbf{ Parity reference $b_{3}$ Number of Cycles 32 of 8 Elements}

 (  127  192  95  240  136  51  170  0  ) (  126  193  94  241  137  50  171  1  ) (  124  194  92  242  139  49  169  2  ) (  125  195  93  243  138  48  168  3  )

 (  121  197  88  244  142  54  173  4  ) (  120  196  89  245  143  55  172  5  ) (  122  199  91  246  141  52  174  6  ) (  123  198  90  247  140  53  175  7  )

 (  115  202  80  248  132  57  165  8  ) (  114  203  81  249  133  56  164  9  ) (  112  200  83  250  135  59  166  10  ) (  113  201  82  251  134  58  167  11  )

 (  117  207  87  252  130  60  162  12  ) (  116  206  86  253  131  61  163  13  ) (  118  205  84  254  129  62  161  14  ) (  119  204  85  255  128  63  160  15  )

 (  152  212  190  225  110  38  74  16  ) (  153  213  191  224  111  39  75  17  ) (  155  214  189  227  109  36  73  18  ) (  154  215  188  226  108  37  72  19  )

 (  158  209  185  229  104  35  77  20  ) (  159  208  184  228  105  34  76  21  ) (  157  211  186  231  107  33  78  22  ) (  156  210  187  230  106  32  79  23  )

 (  148  222  177  233  98  44  69  24  ) (  149  223  176  232  99  45  68  25  ) (  151  220  178  235  97  46  70  26  ) (  150  221  179  234  96  47  71  27  )

 (  146  219  182  237  100  41  66  28  ) (  147  218  183  236  101  40  67  29  ) (  145  217  181  239  103  43  65  30  ) (  144  216  180  238  102  42  64  31  )

 \textbf{ Parity reference $b_{4}$ Number of Cycles 32 of 8 Elements}

 (  127  192  95  15  119  204  85  0  ) (  126  193  94  14  118  205  84  1  ) (  124  194  92  13  116  206  86  2  ) (  125  195  93  12  117  207  87  3  )

 (  121  197  88  11  113  201  82  4  ) (  120  196  89  10  112  200  83  5  ) (  122  199  91  9  114  203  81  6  ) (  123  198  90  8  115  202  80  7  )

 (  103  212  65  30  110  217  74  16  ) (  102  213  64  31  111  216  75  17  ) (  100  214  66  28  109  219  73  18  ) (  101  215  67  29  108  218  72  19  )

 (  97  209  70  26  104  220  77  20  ) (  96  208  71  27  105  221  76  21  ) (  98  211  69  24  107  222  78  22  ) (  99  210  68  25  106  223  79  23  )

 (  176  232  156  45  187  230  149  32  ) (  177  233  157  44  186  231  148  33  ) (  179  234  159  47  184  228  150  34  ) (  178  235  158  46  185  229  151  35  )

 (  182  237  155  41  189  227  146  36  ) (  183  236  154  40  188  226  147  37  ) (  181  239  152  43  190  225  145  38  ) (  180  238  153  42  191  224  144  39  )

 (  168  252  130  60  162  243  138  48  ) (  169  253  131  61  163  242  139  49  ) (  171  254  129  62  161  241  137  50  ) (  170  255  128  63  160  240  136  51  )

 (  174  249  133  56  164  246  141  52  ) (  175  248  132  57  165  247  140  53  ) (  173  251  134  58  167  244  142  54  ) (  172  250  135  59  166  245  143  55  )

 \textbf{ Parity reference $b_{5}$ Number of Cycles 32 of 8 Elements}

 (  127  192  160  15  119  204  170  0  ) (  126  193  161  14  118  205  171  1  ) (  124  194  163  13  116  206  169  2  ) (  125  195  162  12  117  207  168  3  )

 (  121  197  167  11  113  201  173  4  ) (  120  196  166  10  112  200  172  5  ) (  122  199  164  9  114  203  174  6  ) (  123  198  165  8  115  202  175  7  )

 (  103  212  190  30  110  217  181  16  ) (  102  213  191  31  111  216  180  17  ) (  100  214  189  28  109  219  182  18  ) (  101  215  188  29  108  218  183  19  )

 (  97  209  185  26  104  220  178  20  ) (  96  208  184  27  105  221  179  21  ) (  98  211  186  24  107  222  177  22  ) (  99  210  187  25  106  223  176  23  )

 (  79  232  156  45  68  230  149  32  ) (  78  233  157  44  69  231  148  33  ) (  76  234  159  47  71  228  150  34  ) (  77  235  158  46  70  229  151  35  )

 (  73  237  155  41  66  227  146  36  ) (  72  236  154  40  67  226  147  37  ) (  74  239  152  43  65  225  145  38  ) (  75  238  153  42  64  224  144  39  )

 (  87  252  130  60  93  243  138  48  ) (  86  253  131  61  92  242  139  49  ) (  84  254  129  62  94  241  137  50  ) (  85  255  128  63  95  240  136  51  )

 (  81  249  133  56  91  246  141  52  ) (  80  248  132  57  90  247  140  53  ) (  82  251  134  58  88  244  142  54  ) (  83  250  135  59  89  245  143  55  )

 \textbf{ Parity reference $b_{6}$ Number of Cycles 32 of 8 Elements}

 (  127  63  95  15  119  51  85  0  ) (  126  62  94  14  118  50  84  1  ) (  124  61  92  13  116  49  86  2  ) (  125  60  93  12  117  48  87  3  )

 (  121  58  88  11  113  54  82  4  ) (  120  59  89  10  112  55  83  5  ) (  122  56  91  9  114  52  81  6  ) (  123  57  90  8  115  53  80  7  )

 (  103  43  65  30  110  38  74  16  ) (  102  42  64  31  111  39  75  17  ) (  100  41  66  28  109  36  73  18  ) (  101  40  67  29  108  37  72  19  )

 (  97  46  70  26  104  35  77  20  ) (  96  47  71  27  105  34  76  21  ) (  98  44  69  24  107  33  78  22  ) (  99  45  68  25  106  32  79  23  )

 (  192  160  240  136  204  170  255  128  ) (  193  161  241  137  205  171  254  129  ) (  195  162  243  138  207  168  252  130  ) (  194  163  242  139  206  169  253  131  )

 (  198  165  247  140  202  175  248  132  ) (  199  164  246  141  203  174  249  133  ) (  197  167  244  142  201  173  251  134  ) (  196  166  245  143  200  172  250  135  )

 (  216  180  238  153  213  191  224  144  ) (  217  181  239  152  212  190  225  145  ) (  219  182  237  155  214  189  227  146  ) (  218  183  236  154  215  188  226  147  )

 (  222  177  233  157  211  186  231  148  ) (  223  176  232  156  210  187  230  149  ) (  221  179  234  159  208  184  228  150  ) (  220  178  235  158  209  185  229  151  )

 \textbf{ Parity reference $b_{7}$ Number of Cycles 32 of 8 Elements}

========================================================

\end{center}
\end{tiny}
 \center \textbf{Table 1:} \\
\begin{small}\center \textbf{ Elements and cycles for string lengths one through eight bits in the encode direction.}

\end{small}
\end{document}